\documentclass{aastex631}

\usepackage{savesym}
\savesymbol{tablenum}
\usepackage{siunitx}
\restoresymbol{SIX}{tablenum}

\usepackage{graphicx,epsf}
\usepackage{subfigure}
\usepackage{graphicx}	
\usepackage{amsmath}	
\usepackage{amssymb}	
\usepackage{newtxtext,newtxmath}
\usepackage{siunitx}
\usepackage{footnote}
\usepackage{slantsc}

\def\one{\textts{{\,\textsc{i}}}}
\def\two{\textts{ {\,\textsc{ii}}}}
\def\three{\textts{ {\,\textsc{iii}}}}
\def\four{\textts{ {\,\textsc{iv}}}}
\def\five{\textts{ {\,\textsc{v}}} }
\def\six{\textts{ {\,\textsc{vi}}} }

\DeclareRobustCommand{\ion}[2]{\relax\ifmmode\ifx\testbx\f@series{\mathbf{#1\,\mathsc{#2}}}\else{\mathrm{#1\,\mathsc{#2}}}\fi\else\textup{#1\,{\mdseries\textsc{#2}}}\fi}

\begin{document}

\title{An Updated Detection Pipeline for Precursor Emission in Type II Supernova 2020tlf}

\correspondingauthor{Wynn Jacobson-Gal\'{a}n (he, him, his)}
\email{wynnjg@caltech.edu}

\author[0000-0002-3934-2644]{W.~V.~Jacobson-Gal\'{a}n}
\altaffiliation{NASA Hubble Fellow}
\affil{Department of Astronomy and Astrophysics, California Institute of Technology, Pasadena, CA 91125, USA}
\affil{Department of Astronomy, University of California, Berkeley, CA 94720-3411, USA}

\author{S.~Gonzalez}
\affil{Department of Astronomy, University of California, Berkeley, CA 94720-3411, USA}

\author{S.~Patel}
\affil{Department of Astronomy, University of California, Berkeley, CA 94720-3411, USA}

\author[0000-0003-0599-8407]{L.~Dessart}
\affil{Institut d’Astrophysique de Paris, CNRS-Sorbonne Université, 98 bis boulevard Arago, F-75014 Paris, France}

\author[0000-0002-6230-0151]{D.~O.~Jones}
\affiliation{Institute for Astronomy, University of Hawai’i, 640 N. A’ohoku Pl., Hilo, HI 96720, USA}


\author[0000-0001-5126-6237]{D.~L.~Coppejans}
\affiliation{Department of Physics, University of Warwick, Coventry CV4 7AL, United Kingdom}

\author[0000-0001-9494-179X]{G.~Dimitriadis}
\affiliation{School of Physics, Trinity College Dublin, The University of Dublin, Dublin, Ireland}

\author[0000-0002-2445-5275]{R.~J.~Foley}
\affiliation{Department of Astronomy and Astrophysics, University of California, Santa Cruz, CA 95064, USA}

\author[0000-0002-5740-7747]{C.~D.~Kilpatrick}
\affil{Center for Interdisciplinary Exploration and Research in Astrophysics (CIERA), and Department of Physics and Astronomy, Northwestern University, Evanston, IL 60208, USA}

\author[0000-0002-4513-3849]{D.~J.~Matthews}
\affiliation{Department of Physics, Illinois Institute of Technology, Chicago IL 60616 USA}

\author[0000-0002-3825-0553]{S.~Rest}
\affiliation{Department of Physics and Astronomy, The Johns Hopkins University, Baltimore, MD 21218, USA}

\author[0000-0003-0794-5982]{G.~Terreran}
\affil{Las Cumbres Observatory, 6740 Cortona Dr, Suite 102, Goleta, CA 93117-5575, USA}

\author[0000-0002-6298-1663]{P.~D.~Aleo}
\affil{Department of Astronomy, University of Illinois at Urbana-Champaign, 1002 W. Green St., IL 61801, USA}
\affil{Center for Astrophysical Surveys, National Center for Supercomputing Applications, Urbana, IL, 61801, USA}

\author[0000-0002-4449-9152]{K.~Auchettl}
\affil{Department of Astronomy and Astrophysics, University of California, Santa Cruz, CA 95064, USA}
\affil{School of Physics, The University of Melbourne, VIC 3010, Australia}

\author[0000-0003-0526-2248]{P.~K.~Blanchard}
\affiliation{Center for Interdisciplinary Exploration and Research in Astrophysics (CIERA), and Department of Physics and Astronomy, Northwestern University, Evanston, IL 60208, USA}

\author[0000-0003-4263-2228]{D.~A.~Coulter}
\affil{Space Telescope Science Institute, Baltimore, MD 21218}

\author[0000-0002-5680-4660]{K.~W.~Davis}
\affil{Department of Astronomy and Astrophysics, University of California, Santa Cruz, CA 95064, USA}

\author[0000-0001-5486-2747]{T.~J.~L.~de Boer}
\affil{Institute for Astronomy, University of Hawaii, 2680 Woodlawn Drive, Honolulu, HI 96822, USA}

\author[0000-0003-4587-2366]{L.~DeMarchi}
\affiliation{Center for Interdisciplinary Exploration and Research in Astrophysics (CIERA), and Department of Physics and Astronomy, Northwestern University, Evanston, IL 60208, USA}

\author[0000-0001-7081-0082]{M.~R.~Drout}
\affiliation{David A. Dunlap Department of Astronomy and Astrophysics, University of Toronto, 50 St. George Street, Toronto, Ontario, M5S 3H4, Canada}

\author[0000-0003-1714-7415]{N.~Earl}
\affil{Department of Astronomy, University of Illinois at Urbana-Champaign, 1002 W. Green St., IL 61801, USA}

\author[0000-0003-4906-8447]{A.~Gagliano}
\affil{The NSF AI Institute for Artificial Intelligence and Fundamental Interactions}

\author[0000-0002-8526-3963]{C.~Gall}
\affil{DARK, Niels Bohr Institute, University of Copenhagen, Jagtvej 128, 2200 Copenhagen, Denmark}

\author[0000-0002-4571-2306]{J.~Hjorth}
\affil{DARK, Niels Bohr Institute, University of Copenhagen, Jagtvej 128, 2200 Copenhagen, Denmark}

\author[0000-0003-1059-9603]{M.~E.~Huber}
\affil{Institute for Astronomy, University of Hawaii, 2680 Woodlawn Drive, Honolulu, HI 96822, USA}

\author[0000-0003-2405-2967]{A.~L.~Ibik}
\affiliation{David A. Dunlap Department of Astronomy and Astrophysics, University of Toronto, 50 St. George Street, Toronto, Ontario, M5S 3H4, Canada}

\author[0000-0002-0763-3885]{D. Milisavljevic}
\affil{Department of Physics and Astronomy, Purdue University, 525 Northwestern Avenue, West Lafayette, IN 47907, USA}

\author[0000-0001-8415-6720]{Y.-C.~Pan}
\affil{Graduate Institute of Astronomy, National Central University, 300 Zhongda Road, Zhongli, Taoyuan 32001, Taiwan}

\author[0000-0002-4410-5387]{A.~Rest}
\affil{Space Telescope Science Institute, Baltimore, MD 21218}
\affiliation{Department of Physics and Astronomy, The Johns Hopkins University, Baltimore, MD 21218, USA}

\author[0000-0003-1724-2885]{R.~Ridden-Harper}
\affiliation{School of Physical and Chemical Sciences | Te Kura Mat\={u}, University of Canterbury, Private Bag 4800, Christchurch 8140, New Zealand}

\author[0000-0002-7559-315X]{C.~Rojas-Bravo}
\affil{Department of Astronomy and Astrophysics, University of California, Santa Cruz, CA 95064,
USA}

\author[0000-0003-2445-3891]{M.~R.~Siebert}
\affil{Space Telescope Science Institute, Baltimore, MD 21218}

\author[0000-0001-9535-3199]{K.~W.~Smith}
\affil{Astrophysics Research Centre, School of Mathematics and Physics, Queen’s University Belfast, Belfast BT7 1NN, UK}

\author[0000-0002-5748-4558]{K.~Taggart}
\affil{Department of Astronomy and Astrophysics, University of California, Santa Cruz, CA 95064,
USA}

\author[0000-0002-1481-4676]{S.~Tinyanont}
\affil{Department of Astronomy and Astrophysics, University of California, Santa Cruz, CA 95064,
USA}

\author[0000-0001-5233-6989]{Q.~Wang}
\affiliation{Department of Physics and Kavli Institute for Astrophysics and Space Research, Massachusetts Institute of Technology, Cambridge, MA 02139, USA}

\author[0000-0002-0632-8897]{Y.~Zenati}
\affiliation{Department of Physics and Astronomy, The Johns Hopkins University, Baltimore, MD 21218, USA}
\affil{Space Telescope Science Institute, Baltimore, MD 21218}


\begin{abstract}

We present a new photometric pipeline for the detection of pre-supernova (pre-SN) emission in the Young Supernova Experiment (YSE) sky survey. The method described is applied to SN~2020tlf, a type II SN (SN II) with precursor emission in the last $\sim$100~days before first light. We re-analyze the YSE $griz$-band light curves of SN~2020tlf and provide revised pre-explosion photometry that includes a robust list of confident detection and limiting magnitudes. Compared to the results of \cite{wjg22}, this new analysis yields fewer total $r/i/z$-band  pre-SN detections at phases $> -100$~days. Furthermore, we discourage the use of the blackbody modeling of the pre-explosion spectral energy distribution, the pre-SN bolometric light curve and the blackbody model parameters presented in \cite{wjg22}. Nevertheless, binned photometry of SN~2020tlf confirms a consistent progenitor luminosity of $\sim$10$^{40}$~erg~s$^{-1}$ before explosion.\\
\end{abstract}




\section{A Refined Pre-Supernova Detection Algorithm}


Here we present an updated procedure for detecting pre-supernova (SN) emission in Pan-STARRS images obtained through the Young Supernova Experiment (YSE) transient survey and revised pre-SN photometry of SN~2020tlf \citep{wjg22}. This process has been employed successfully in recent and forthcoming publications (e.g., \citealt{Ransome24}; Taggart et al., in prep.). Similar to the procedure described in Section~2.1 of \cite{wjg22}, we construct control light curves of the host-galaxy emission, now placing apertures along an isophotal contour rather than just an isophotal ellipse, avoiding the SN location itself. Photometry is then calculated in each difference image at each background aperture location and the SN position. We then perform an idealized artificial source injection where increasing amounts of discrete flux are added to the photometric measurement of each chosen aperture until the flux to error ratio is greater than three (i.e., a 3-$\sigma$ detection). We use the fraction of apertures where the injected source is recovered at $>$3$-\sigma$ significance to estimate the recovery fraction as a function of the injected source flux. This injection method is consistent with the artificial source injection performed on science images during the {\tt Photpipe} \citep{Rest+05} reduction (see Section~2.1 of \citealt{Ransome24} for a more detailed discussion of the routine). However, the injection of artificial sources during the photometric reduction itself is significantly slower and computationally intensive. Therefore, because it yields similar limits on the pre-SN emission, we adopt the idealized artificial-source injection procedure for our analysis. 

\begin{figure}
\centering
\subfigure{\includegraphics[width=0.49\textwidth]{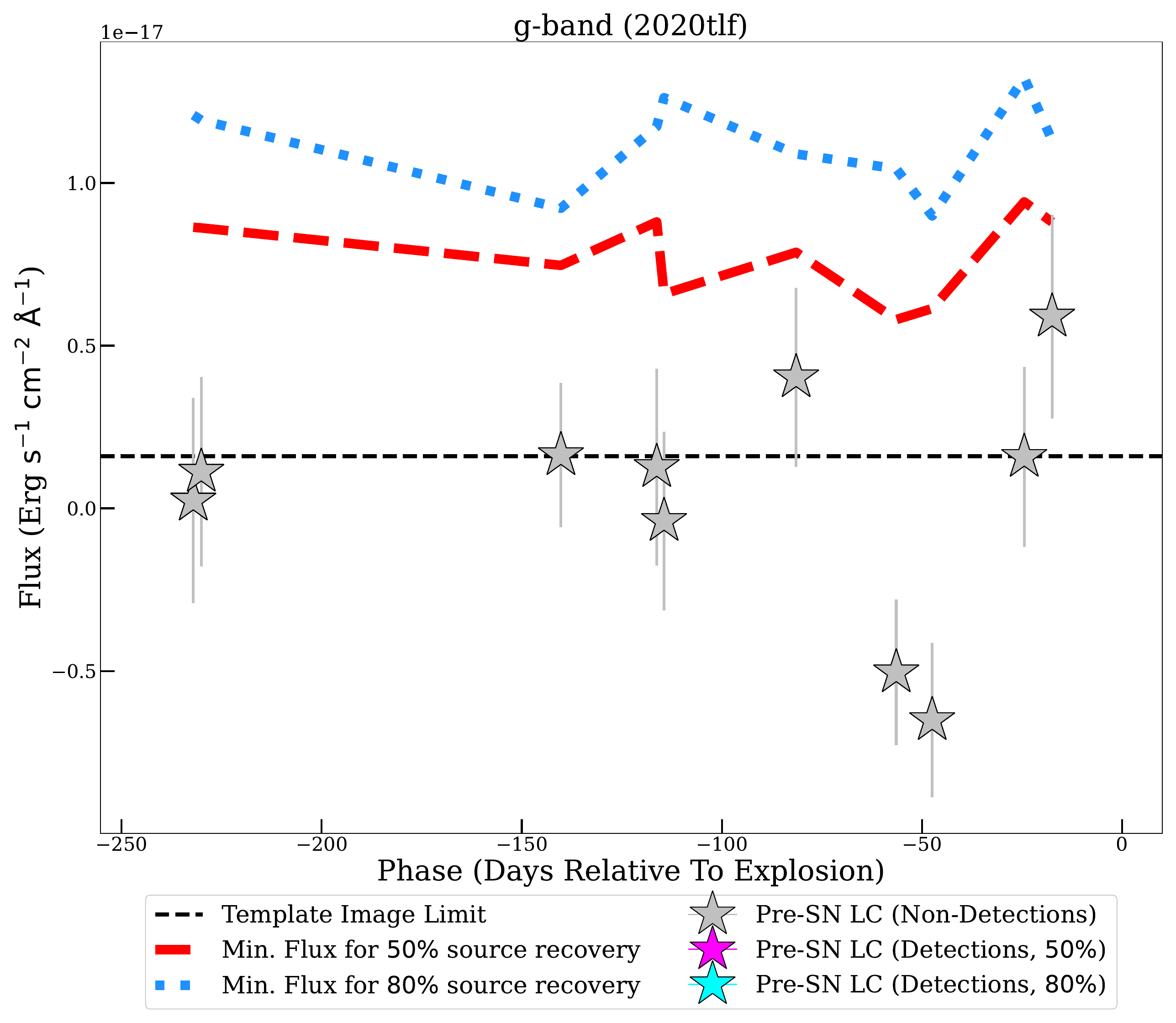}}
\subfigure{\includegraphics[width=0.49\textwidth]{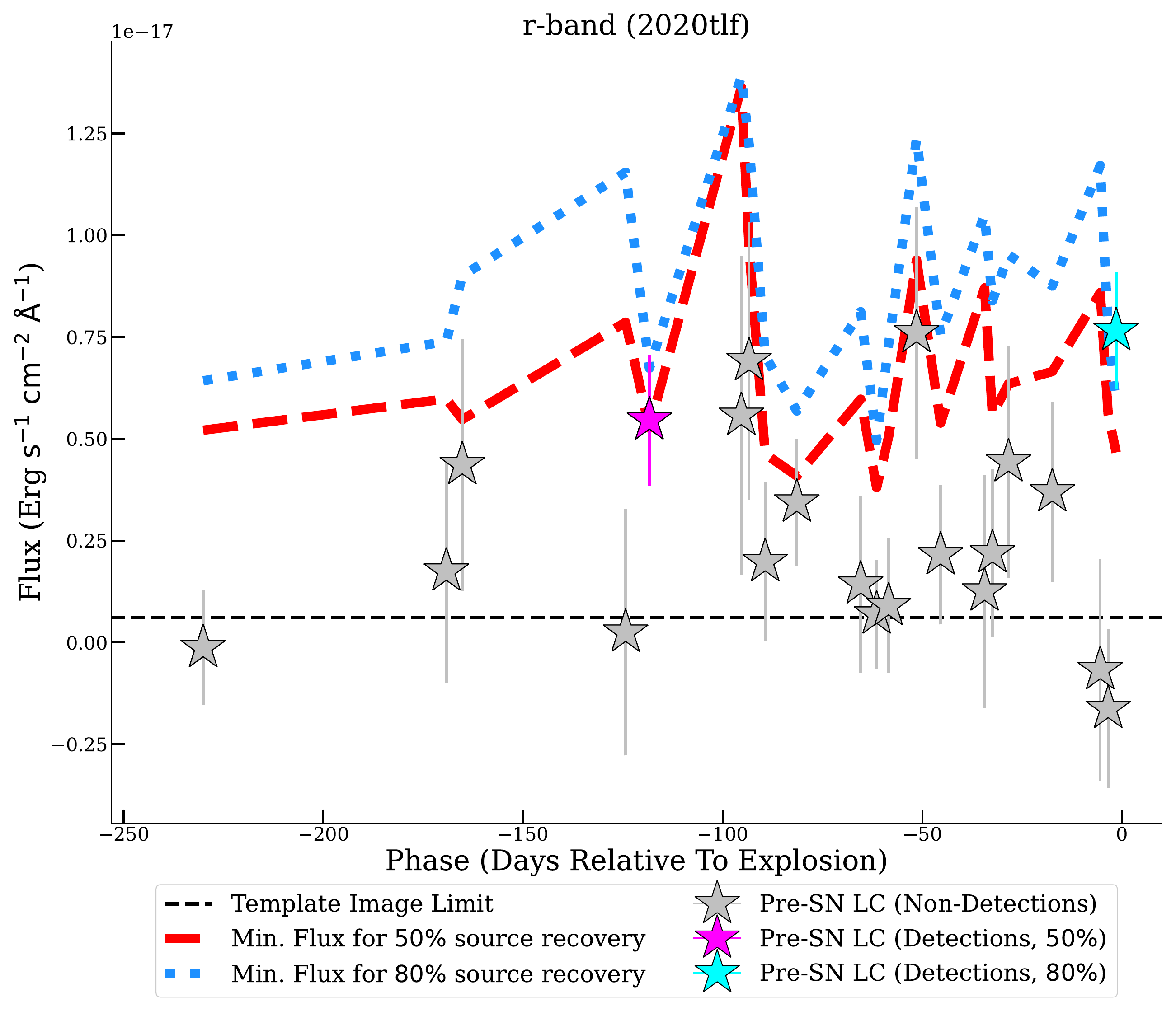}}\\
\subfigure{\includegraphics[width=0.49\textwidth]{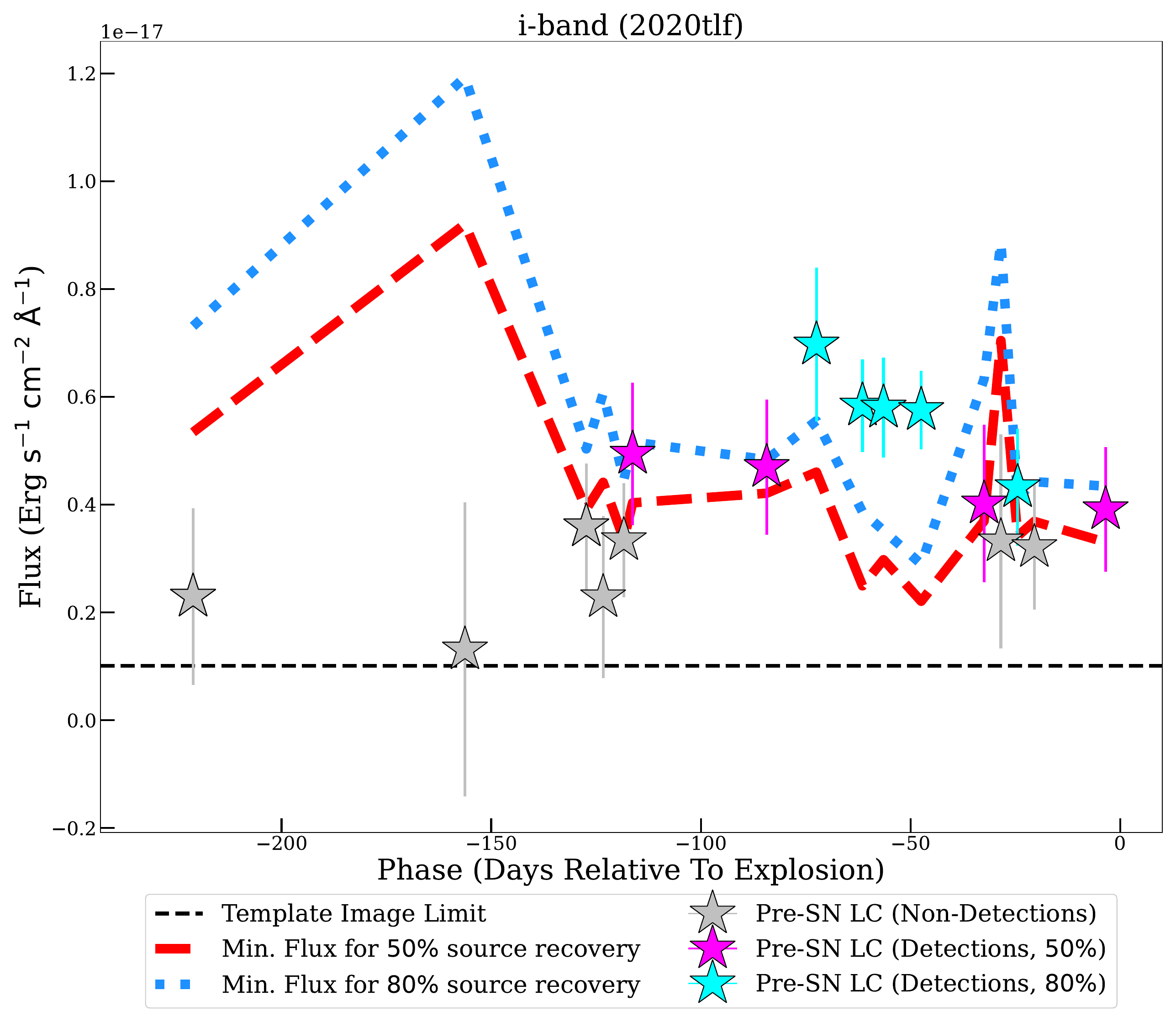}}
\subfigure{\includegraphics[width=0.49\textwidth]{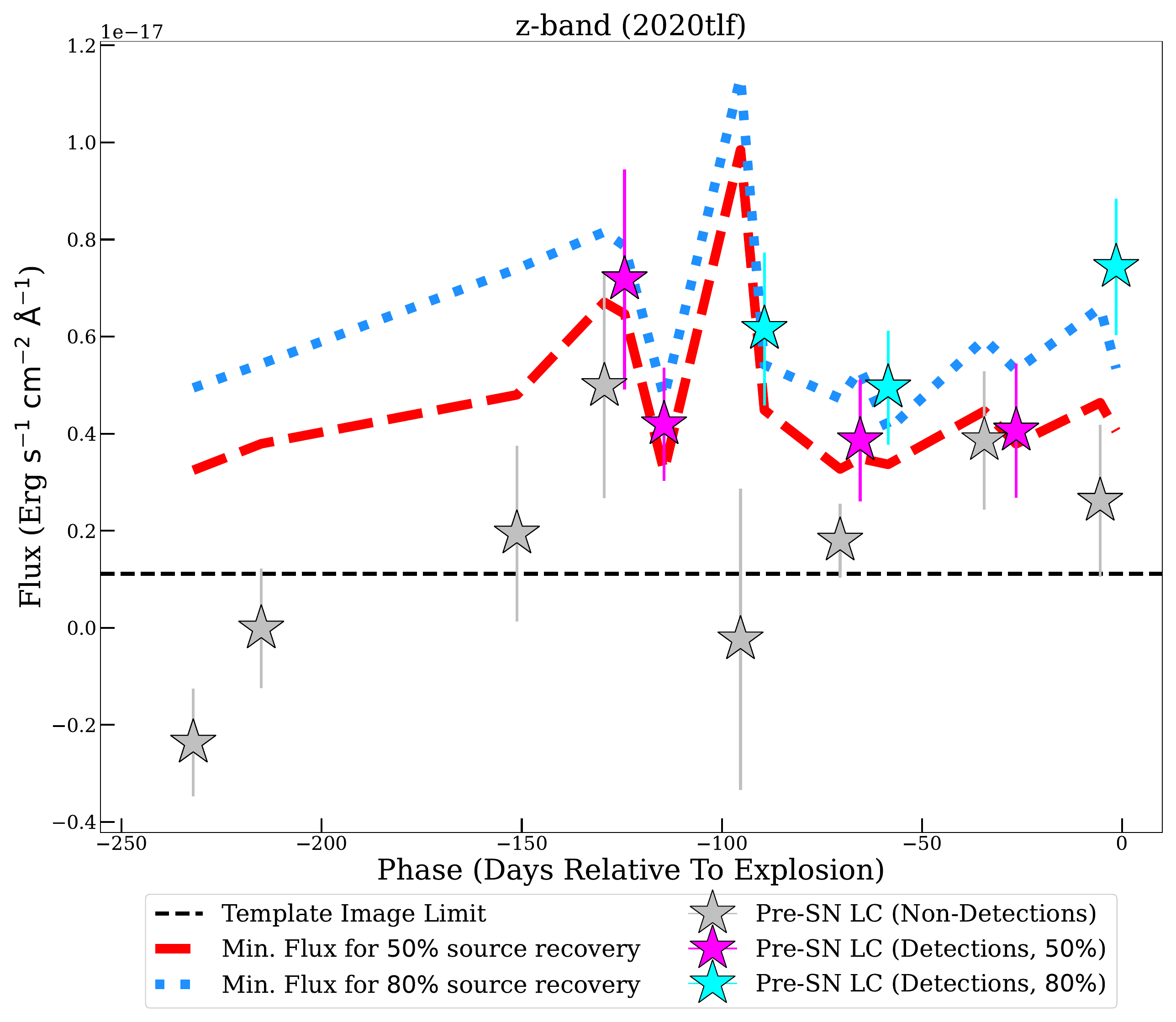}}\\
\subfigure{\includegraphics[width=0.69\textwidth]{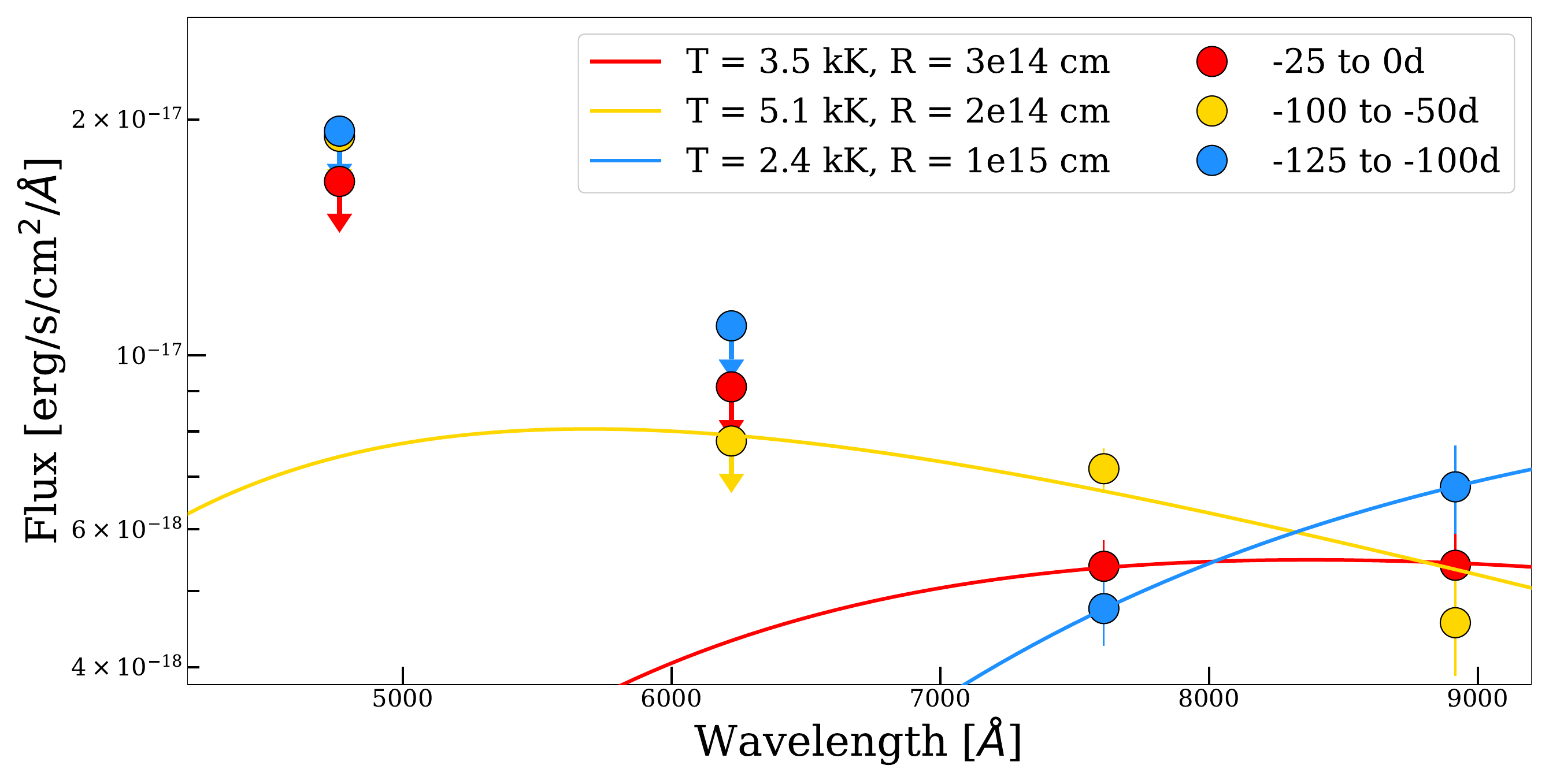}}
\caption{ {\it Top/Middle Panel:} Pre-explosion $griz$-band light curve of SN~2020tlf. Detections above 80\% and 50\% recovery levels shown as cyan and magenta stars, respectively. Calculated flux at the SN location that is below these recovery levels is shown as grey stars. Minimum fluxes for 80\% and 50\% injected source recovery shown as blue dotted and red dashed lines, respectively. Flux measured at SN location from pre-explosion PS1 3$\pi$ survey imaging shown as dashed black lines. {\it Bottom Panel:} Spectral energy distribution using $griz$-band limits and detections in 50~day pre-SN light curve bins. Best-fitting blackbody models shown as solid lines. \label{fig:preSN_LC_nobin} }
\end{figure}

In order to determine if a measurement at the SN location is a significant detection, we compare the derived limiting magnitude to the photometric measurement output from {\tt Photpipe} at the SN location and label a detection as real if the latter is brighter than the former. We calculate the limiting magnitude for an 80\% recovery fraction in the background apertures; i.e., the flux at which $>$80\% of the injected sources are recovered at $>$3$-\sigma$ significance. Based on the analysis of PS1 images presented in \cite{Metcalfe13}, a 50\% efficiency threshold was required for fake sources to be recovered at a 5-$\sigma$ level. However, given that additional uncertainties in the images are not quantified in this threshold, we chose to adopt a more conservative recovery level of 80\% in order to verify true pre-explosion detections. The $griz$ pre-SN light curves of SN~2020tlf, with both detections and fluxes consistent with zero, are shown in Figure \ref{fig:preSN_LC_nobin}. For completeness, we also show pre-SN light curves with 50- and 100-day binning in the Supplementary Figures.\footnote{\url{https://github.com/wynnjacobson-galan/PreSN}} Furthermore, we include $griz$ difference images of the pre-explosion source at $>$50\% and $>$80\% detection levels.  The revised pre-SN photometry is presented in Tables \ref{tab:phot}-\ref{tab:phot2} and is also available online.$^1$



\citet{wjg22} presented 21, 17, and 15 epochs of $riz$ photometry, of which 2, 11, and 6 epochs were considered detections, respectively.  With the new method presented here, 1, 5, and 3 epochs are considered detections at an 80\% recovery level in $riz$, respectively, while 1, 6, and 3 epochs previously considered detections are not deemed significant with the new analysis. Alternatively, with the new method presented, 2, 9, and 7 epochs are considered detections at $>$50\% recovery levels in $riz$, respectively. The photometry derived in this updated method also translates to pre-explosion absolute magnitudes of $M \approx -11.5$ to $-12$~mag in $riz$. However, we note that magnitudes derived are from difference images and that there is some underlying progenitor star flux that has been subtracted off. To quantify this, measure forced photometry at the SN location in the pre-explosion PS1 3$\pi$ survey imaging of the SN site from 28 Feb. 2011 to 21 Feb. 2014 ($\delta t = -3478 ~ {\rm to} ~ -2389$~days before first light). We present these fluxes for each filter in Figure \ref{fig:preSN_LC_nobin}.


Similar to \citet{wjg22}, we apply a blackbody model to the pre-explosion spectral-energy distribution (SED) of SN~2020tlf after 50~day phase binning has been applied to the multi-band light curves. Given the measured flux at the SN location in the template images, the flux removed in difference imaging could be significant enough to impact the estimated blackbody model parameters. Consequently, we add the template image flux to the binned difference flux before fitting a blackbody model and caution that derived parameters such as blackbody radii and luminosities should be treated as conservative upper limits on the true pre-explosion blackbody emission. For the blackbody model fits presented in Figure \ref{fig:preSN_LC_nobin}, the pre-explosion ($\delta t > -100$~days) bolometric luminosity is $\sim$10$^{40}$~erg~s$^{-1}$, in addition to blackbody radii and temperatures of $\sim$~$(2-3)\times 10^{14}$~cm and $\sim$3000--5000~K, all similar to what is presented in Figure~6 of \citet{wjg22}. Interestingly, the blackbody parameters derived from the SED in the $\delta t = -150$ to $-100$~day bin suggest a higher bolometric luminosity ($\sim${}$2\times10^{40}$~erg~s$^{-1}$ and blackbody radius ($\sim$10$^{15}$~cm), but lower temperature ($\sim$2500~K). Nevertheless, similar to the initial analysis, the updated analysis confirms rising pre-explosion flux in SN~2020tlf within the last $\sim$100~days prior to first light. 


\begin{deluxetable}{cccccc}
\tablecaption{Pre-SN PS1/YSE Photometry of SN~2020tlf \label{tab:phot}}
\tablecolumns{6}
\tablewidth{0pt}
\tablehead{\colhead{MJD} & \colhead{Phase$^a$} & \colhead{Filter} & \colhead{Magnitude} & \colhead{Uncertainty} & \colhead{Recovery Efficiency}}
\startdata
58866.7 & -232.1 & $g$ & 21.9 & -- & $<50\%$ \\
58958.5 & -140.2 & $g$ & 22.0 & -- & $<50\%$ \\
58982.5 & -116.3 & $g$ & 21.8 & -- & $<50\%$ \\
58984.3 & -114.4 & $g$ & 22.2 & -- & $<50\%$ \\
59017.3 & -81.4 & $g$ & 22.0 & -- & $<50\%$ \\
59042.3 & -56.4 & $g$ & 22.3 & -- & $<50\%$ \\
59051.3 & -47.5 & $g$ & 22.2 & -- & $<50\%$ \\
59074.3 & -24.5 & $g$ & 21.8 & -- & $<50\%$ \\
59081.2 & -17.5 & $g$ & 21.8 & -- & $<50\%$ \\
58868.7 & -230.1 & $r$ & 21.8 & -- & $<50\%$\\
58929.5 & -169.2 & $r$ & 21.7 & -- & $<50\%$\\
58933.5 & -165.2 & $r$ & 21.8 & -- & $<50\%$\\
58974.4 & -124.3 & $r$ & 21.4 & -- & $<50\%$\\
58980.4 & -118.4 & $r$ & 21.8 & 0.3 & $>50\%$\\
59003.4 & -95.3 & $r$ & 20.8 & -- & $<50\%$\\
59005.3 & -93.4 & $r$ & 21.2 & -- & $<50\%$\\
59009.4 & -89.4 & $r$ & 22.0 & -- & $<50\%$\\
59017.3 & -81.4 & $r$ & 22.1 & -- & $<50\%$\\
59033.3 & -65.5 & $r$ & 21.7 & -- & $<50\%$\\
59037.3 & -61.5 & $r$ & 22.2 & -- & $<50\%$\\
59040.3 & -58.5 & $r$ & 21.9 & -- & $<50\%$\\
59047.3 & -51.5 & $r$ & 21.2 & -- & $<50\%$\\
59053.3 & -45.5 & $r$ & 21.8 & -- & $<50\%$\\
59064.3 & -34.4 & $r$ & 21.3 & -- & $<50\%$\\
59066.3 & -32.5 & $r$ & 21.8 & -- & $<50\%$\\
59070.3 & -28.4 & $r$ & 21.6 & -- & $<50\%$\\
59081.2 & -17.5 & $r$ & 21.6 & -- & $<50\%$\\
59093.2 & -5.5 & $r$ & 21.3 & -- & $<50\%$\\
59095.2 & -3.5 & $r$ & 21.7 & -- & $<50\%$\\
59097.2 & -1.5 & $r$ & 21.4 & 0.2 & $>80\%$\\
\enddata
\tablenotetext{a}{Relative to first light (MJD 59098.74)}
\end{deluxetable}

\begin{deluxetable}{cccccc}
\tablecaption{Pre-SN PS1/YSE Photometry of SN~2020tlf (cont.) \label{tab:phot2}}
\tablecolumns{6}
\tablewidth{0pt}
\tablehead{\colhead{MJD} & \colhead{Phase$^a$} & \colhead{Filter} & \colhead{Magnitude} & \colhead{Uncertainty} & \colhead{Recovery Efficiency}}
\startdata
58877.6 & -221.1 & $i$ & 21.4 & -- & $<50\%$\\
58942.5 & -156.3 & $i$ & 20.8 & -- & $<50\%$\\
58971.4 & -127.3 & $i$ & 21.7 & -- & $<50\%$\\
58975.4 & -123.3 & $i$ & 21.6 & -- & $<50\%$\\
58980.4 & -118.4 & $i$ & 21.9 & -- & $<50\%$\\
58982.5 & -116.3 & $i$ & 21.4 & 0.3 & $>50\%$\\
59014.4 & -84.3 & $i$ & 21.5 & 0.3 & $>50\%$\\
59026.3 & -72.4 & $i$ & 21.1 & 0.2 & $>80\%$\\
59037.3 & -61.5 & $i$ & 21.3 & 0.2 & $>80\%$\\
59042.3 & -56.4 & $i$ & 21.3 & 0.2 & $>80\%$\\
59051.3 & -47.5 & $i$ & 21.3 & 0.1 & $>80\%$\\
59058.3 & -40.4 & $i$ & 21.5 & -- & $<50\%$\\
59066.3 & -32.5 & $i$ & 21.7 & 0.4 & $>50\%$\\
59070.3 & -28.5 & $i$ & 21.1 & -- & $<50\%$\\
59074.3 & -24.5 & $i$ & 21.6 & 0.3 & $>80\%$\\
59078.3 & -20.5 & $i$ & 21.8 & -- & $<50\%$\\
59095.2 & -3.5 & $i$ & 21.7 & 0.3 & $>50\%$\\
58866.7 & -232.1 & $z$ & 21.6 & -- & $<50\%$\\
58883.7 & -215.1 & $z$ & 21.4 & -- & $<50\%$\\
58947.5 & -151.2 & $z$ & 21.1 & -- & $<50\%$\\
58969.4 & -129.4 & $z$ & 20.8 & -- & $<50\%$\\
58974.4 & -124.3 & $z$ & 20.7 & 0.3 & $>50\%$\\
58984.3 & -114.4 & $z$ & 21.3 & 0.3 & $>50\%$\\
59003.4 & -95.3 & $z$ & 20.4 & -- & $<50\%$\\
59009.4 & -89.4 & $z$ & 20.9 & 0.3 & $>80\%$\\
59028.3 & -70.5 & $z$ & 21.6 & -- & $<50\%$\\
59033.3 & -65.5 & $z$ & 21.4 & 0.4 & $>50\%$\\
59040.3 & -58.5 & $z$ & 21.1 & 0.3 & $>80\%$\\
59064.3 & -34.4 & $z$ & 21.2 & -- & $<50\%$\\
59072.3 & -26.5 & $z$ & 21.3 & 0.4 & $>50\%$\\
59093.2 & -5.5 & $z$ & 21.2 & -- & $<50\%$\\
59097.2 & -1.5 & $z$ & 20.7 & 0.2 & $>80\%$\\
\enddata
\tablenotetext{a}{Relative to first light (MJD 59098.74)}
\end{deluxetable}

\bibliographystyle{aasjournal} 
\bibliography{references} 

\end{document}